# Effects of post-deposition annealing on the structure and magnetization of PLD grown Yttrium Iron Garnet films

Ravinder Kumar, Z. Hossain and R. C. Budhani*

*Condensed Matter Low-Dimensional Systems Laboratory, Department of Physics, Indian Institute of Technology (IIT) Kanpur – 208016, India.*

**ABSTRACT:**

We report on the recrystallization of 200 nm thick as-grown Yttrium Iron Garnet ($Y_{3.4}Fe_{4.6}O_{12}$) films on (111) face of Gadolinium Gallium Garnet (GGG) single crystals by post-deposition annealing. Epitaxial conversion of the as-grown microcrystalline YIG films was seen after annealing at 800ºC for more than 30 minutes both in ambient oxygen as well as in air. The as-grown oxygen annealed samples at 800ºC for 60 minutes crystallize epitaxially and show excellent figure-of-merit for saturation magnetization ($M_S$ = 3.3 $\mu_B$/f.u., comparable to bulk value) and coercivity ($H_C$ ~ 1.1 Oe). The ambient air annealing at 800ºC with a very slow rate of cooling (2ºC/min) results in a double layer structure with a thicker unstrained epitaxial top layer having the $M_S$ and $H_C$ of 2.9 $\mu_B$/f.u. and 0.12 Oe respectively. The symmetric and asymmetric Reciprocal space maps of both the samples reveal a locking of the in-plane lattice of the film to the in-plane lattice of substrate, indicating a pseudomorphic growth. The residual stress calculated by $\sin^2\psi$ technique is compressive in nature. The lower layer in air annealed sample is highly strained, whereas, the top layer has negligible compressive stress.

## INTRODUCTION:

The development of Spintronics into a viable technology requires novel magnetic materials and ferrimagnetic garnets are one such class due to their very low spin wave damping. Yttrium Iron Garnet (YIG) belongs to the garnet family with native structure $A_3B_2(C.O_4)_3$, where, B – octahedral, C – tetrahedral sites are occupied by Iron (III) ions and Yttrium (III) ions occupy the A – dodecahedral site. The garnet $Y_3Fe_2(Fe.O_4)_3$ (YIG) is known for its high Curie temperature (~ 560 K) and its ferrimagnetic behaviour comes from the opposite iron spin orientation at the B and C sites.[1] Since the discovery of this compound in 1956,[2,3] it remains of considerable scientific interest due to many remarkable properties that it possesses like low spin wave damping (SWD), Faraday rotation (FR), high transmittance for infrared radiation etc., which make it suitable for microwave device applications, magneto-optical (M-O) recording, magnonics, spintronics and caloritronics.[4-13] These properties of YIG can be tuned further by changing the elemental composition, doping, epitaxial orientation, stress engineering and other manipulations.[6,14-20] Since for many applications, YIG is required in a thin film form, the growth of high quality epitaxial films of this material has emerged as a major field of research. It is now well established that gadolinium-gallium-garnet (GGG) [space group $Ia\bar{3}d(O_h^{10})$, lattice parameter = 1.2383 nm] is the best substrate for growth of epitaxial YIG films. The parent garnet belongs to the cubic centrosymmetric space group $Ia\bar{3}d(O_h^{10})$,[1] whereas YIG thin films on GGG show tetragonal distortion which may be caused by growth anisotropy and lattice mismatch between the film and substrate. The pulsed laser deposition technique has been used successfully to deposit high quality films of ferrite materials. Generally, the growth of well crystallized films with good ferrimagnetic ordering of $Y_3Fe_2(Fe.O_4)_3$ on lattice matched and other single crystal substrates requires high temperature annealing after deposition.[6,14,21,22] There are some reports on epitaxial growth of as-grown YIG films by pulsed laser deposition (PLD) but the coercivity of such samples is on higher side.[23,24] Whereas, post-annealing process is helpful to get very low coercivity, which is desirable in spintronics.

*e-mail: rcb@iitk.ac.in

In this work, we investigate the effect of post-deposition annealing (in oxygen and air) on the structure and magnetic properties of 200 nm thick YIG films fabricated using pulsed laser ablation on GGG single crystal substrates with (111) orientation. The annealing at 800°C in full oxygen environment converts the as-deposited, poorly crystalline YIG films to epitaxial single crystal layers. Whereas, a double layer structure formation takes place in films annealed at 800°C in air. The reciprocal space mapping in both the post-treated samples in symmetric (444) as well as asymmetric (642) directions show pseudomorphic growth. The film tries to mimic the substrate lattice and hence shows tetragonal distortion which further results into a structure under stress. The residual stress measured using $\sin^2\psi$ technique is compressive in the oxygen annealed sample and preferentially in the lower layer of air annealed sample. The upper layer in air annealed sample is however fully relaxed. Furthermore, the magnetization measurement for oxygen annealed and air annealed samples give $M_S$ of 3.3 μB/f.u. and 2.9 μB/f.u. and a reasonably low coercivity of 1.1 Oe and 0.12 Oe respectively. The stress-induced magnetic anisotropy constant and magnetic anisotropy field were also calculated for both the samples.

**EXPERIMENTAL METHOD:**

A hard ceramic YIG target was prepared by solid state synthesis with initial powder reactants $Y_2O_3$ and $Fe_2O_3$, mixed in proper stoichiometric proportion. The mixture was first ground and annealed repeatedly between 700°C to 1200°C to get small sized particles of YIG to produce a non-porous target. This powder was compacted to a 2.2 cm diameter pellet by applying 100 MPa pressure and then annealed at 1400°C for 20 hrs. Smaller pellets were also prepared under identical conditions to confirm the elemental composition and homogeneity of YIG target using energy-dispersive X-ray spectroscopy (EDXS) and elemental mapping respectively. Powder X-ray diffractometry was performed to characterize the crystalline structure of prepared target using PANalytical X'Pert PRO four circle diffractometer equipped with Cu-$K_{\alpha 1}$ source (λ = 1.54059 Å). Room temperature Vibrating Sample Magnetometry (VSM) measurements were performed on three different phases of the bulk YIG as well as thin films by using a Physical Property Measurement System (PPMS).

The YIG films were grown by PLD using a KrF Excimer laser (Lambda Physik COMPex Pro, $\lambda$ = 248 nm) of 20 ns pulse width at $800^oC$ in $4.0 \times 10^{-2}$ mbar Oxygen pressure. The laser was fired at a repetition frequency of 10Hz with areal energy of 21.2 kJ/m$^2$ on target surface placed at 50 mm away from the substrate. Typically ~ 200 nm thick films were deposited on $3 \times 3$ mm$^2$ GGG substrates at a growth rate of ~ 0.074 nm/s. The as-grown microcrystalline films were post-annealed to recrystallize via oxygen and air annealing. In the first annealing approach, the samples were annealed at 800°C for 30 and 60 minutes in presence of full Oxygen atmosphere (1 atm Oxygen) just after deposition, followed by cooling to room temperature at a rate of 10°C/min. In the second annealing approach, as-grown films were subjected to annealing at 800°C for different time periods such as 30, 60, 120 and 240 minutes in air. The YIG/GGG(111) samples annealed at $800^oC$ for 30 and 60 minutes in ambient air were cooled down at a rate of 10°C/min, whereas the samples annealed for 120 and 240 minutes were cooled down at a rate of 2°C/min. Elemental composition and homogeneity of all these films were quantified by performing EDXS and elemental mapping respectively. Reciprocal space mapping was done on both types of films by using the PANalytical X'Pert PRO. Pole-figures were also drawn to see the film texture. The residual stress was determined with basic equation derived from the theory of X-ray diffraction.[25-27] The measurement procedure involved identification of different (hkl) planes such as (444), (422), (642), (640), (420) and (400) by using ψ direction tilt at a fix ϕ. Then the 2θ of a particular plane was set and a slow θ-2θ scan was done to get the corresponding diffraction plane peak.

**RESULTS AND DISCUSSION:**

Figure 1 shows the X-ray diffraction profile of powder YIG and subsequent Rietveld refinement analysis of the data. A perfect matching of the sample profile with standard fitting shows that we have a phase pure target with no detectable deviations from the proper stoichiometry. The average elemental composition of the target as determined by EDXS is $Y_{2.9}Fe_{5.2}O_{12}$. Figure 2(a) and 2(b) show the field emission scanning electron microscopy (FESEM) images of the YIG target annealed at 1000°C and 1400°C for 12 hrs respectively. The sample annealed at 1000°C is porous in nature while after annealing at 1400°C, it becomes compact. Further, energy dispersive X-ray elemental mapping revealed homogeneity of all the elements in the target.

*e-mail: rcb@iitk.ac.in

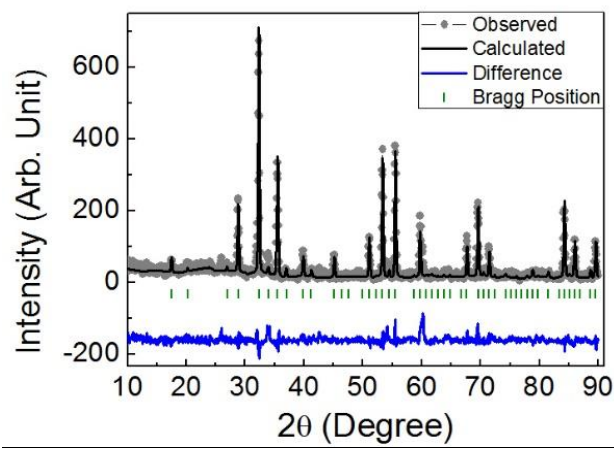

Figure 1: Powder X-ray Diffractogram of the YIG target annealed at 1400°C. Rietveld refinement was done by using fullprof suite (crystallographic tools for Rietveld, profile matching and integrated intensity refinements of X-ray and/or neutron data) shows polycrystalline nature of the material.

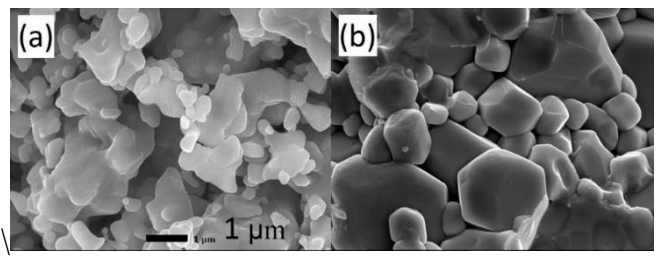

Figure 2: FESEM images of the YIG target; (a) Solid YIG target annealed at 1000°C for 12 hrs. (b) Solid YIG target annealed at 1400°C for 12 hrs. The YIG target annealed at 1000°C is relatively porous in nature compare to the one annealed at 1400°C.

The as-grown films appear dark brown presumably due to oxygen deficiency and have a poor crystalline structure. After post-annealing in oxygen or air, these films recrystallize and appear light yellow. Earlier studies have also shown that the as-grown films deposited at 800°C were annealed in oxygen atmosphere to convert these into epitaxial layers with enhanced magnetic properties.[6,14,21,22] Further, post annealing in ambient air has also been tried successfully to recrystallize as-grown films to yield a low magnetic damping material useful for device fabrication.

Figure (3) shows the X-ray diffraction pattern of various YIG films deposited on (111) surface of GGG and annealed in oxygen and ambient air. In Figure 3(a) we note that the as-grown film has no discernible YIG peaks. The θ-2θ profile only shows the (444) reflection of the substrate with slight asymmetry toward right, suggesting a poorly developed YIG phase. After annealing in oxygen at 800°C for 30 minutes, additional features appear on both sides of the substrate peak. However, after 60 minutes of annealing, the YIG (444) peak with Laue oscillations indicating epitaxial growth emerges at the left side of the GGG (444) peak. We name the YIG sample annealed at 800°C for 60 minutes in full oxygen as "Sample A" and tag the film peak as "A1" (see figure 3(a)). The average elemental composition of the YIG film annealed at 800°C for 60 minutes in full oxygen, quantified by the EDXS analysis is $Y_{3.4}Fe_{4.6}O_{12}$.

In the second annealing approach, the samples were removed from deposition chamber and annealed in ambient air at 800°C for time intervals of 30, 60, 120 and 240 minutes. The X-ray diffraction profile of these samples are shown in figure 3(b). After annealing in air at 800°C for 30 minutes, the same features can be spotted on both sides of the substrate peak, as seen in 30 mins oxygen annealed sample. The ambient air annealing for 60 minutes results in a broad YIG (444) peak with no Laue oscillations. Further increment in annealing time to 120 minutes, yields another film peak at the right side of the substrate peak. These features are enhanced further on annealing for 240 minutes.

We have labelled this as "sample F" and have marked the film peak at left and right to the substrate peak as "F1" and "F2" respectively (see figure 3(b)). In order to find out the origin of these two peaks, we have calculated the d-spacing corresponding to F1 and F2, which are 0.181 nm and 0.178 nm respectively. It is important to note that the d-spacing for F2 peak is close to the $d_{444}$ of bulk YIG (0.179 nm). This observation suggest that the air annealed film has a bilayer structure whose top layer is fully relaxed with d-spacing that of bulk YIG and the bottom layer is highly strained by the substrate thus contributing to peak F1. A similar observation has been made by Mino et al.[28] in cerium doped YIG films deposited on GGG(111) using the conventional RF sputtering technique. They found that the upper layer has almost cubic structure while the lower layer is largely distorted. Eva et al.[29] have also grown YIG on GGG substrates and the films were annealed at 800°C in 1.5 mbar $O_2$ atmosphere followed by slow cooling (1°C/min). They observed decay in the Magneto-Optic Polar Kerr rotation (PKR) amplitudes and attributed it to interface effects. They employed a model of YIG/GGG interface layer thickness but with an opposite sign of the YIG off-diagonal permittivity tensor element (ref. 29). Further, the surface effects in chemical vapour deposited (CVD) YIG films on GGG substrates were investigated by Ramer and Wilts.[30] They used spin-wave resonance method on YIG films (~ 500nm), before and after annealed in dry $O_2$ at 900°C. Their results showed that the magnetic properties for the regions at the air/YIG and YIG/GGG interfaces are different from those of the bulk. In particular, the YIG/GGG interface region thickness increases with annealing. The effect was explained by the diffusion of $Gd^{3+}$ and $Ga^{3+}$ ions into YIG films and hence may produce layers with compensation points.[31-34] In general, the post-deposition annealing treatment at 800°C may cause moderate migration of $Fe^{3+}$ from the YIG film and $Ga^{3+}$ from the GGG substrate across the interface.

*e-mail: rcb@iitk.ac.in

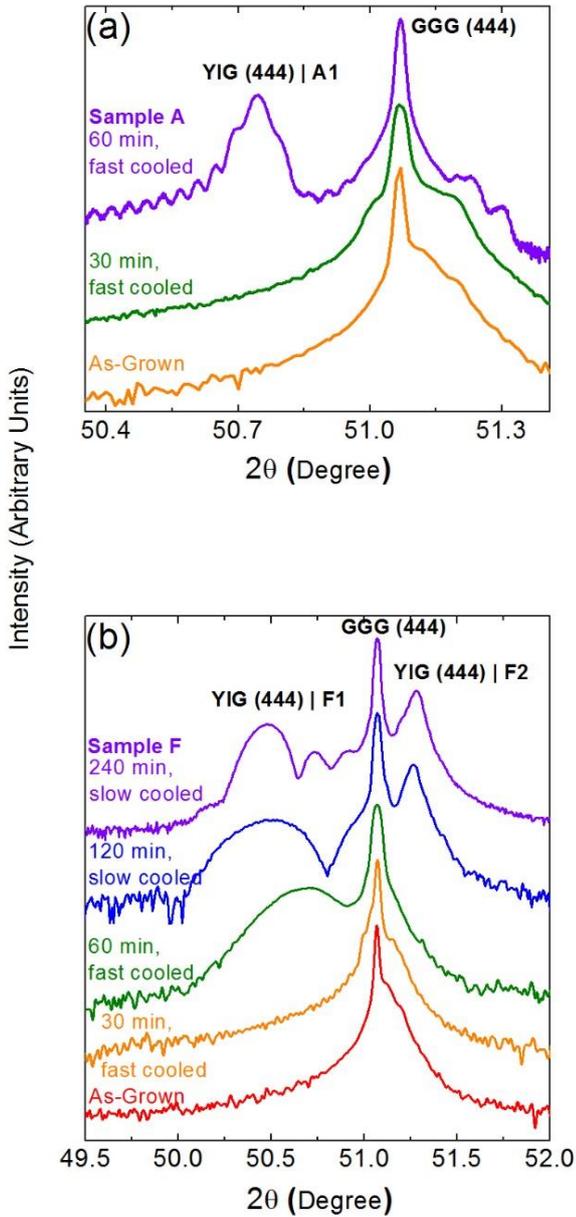

Figure 3: X-ray Diffractogram of 200 nm YIG film on GGG (111) substrate shows recrystallization after post annealing; (a) Samples annealed at 800°C in full oxygen for different annealing time intervals. (b) Samples annealed at 800°C in air for different annealing time intervals with two cooling protocols.

Further, the X-ray diffraction ω scans were taken on sample A and Sample F (not shown here). The full width at half maximum (FWHM) of ω comes out to be 0.029°, 0.067° and 0.043° for Sample A, sample F's F1 and F2 peaks respectively. These low values of the full width at half maximum (FWHM) suggest good crystallinity of A1, F1 and F2 layers.

Reciprocal space map of (444) symmetric direction of sample A and Sample F are shown in Figure 4(a) and 4(b) respectively. Similar maps for both the post-treated samples in (642) asymmetric direction are shown in fig. 4(c) and 4(d) respectively. Reciprocal space maps give clear indication of epitaxial crystallization in both the samples. The in-plane lattice of the film layer is locked to the in-plane lattice of the substrate. This film-substrate in-plane lattice locking represents

*e-mail: rcb@iitk.ac.in

pseudomorphic cubic symmetry and hence results in large stress in the film.

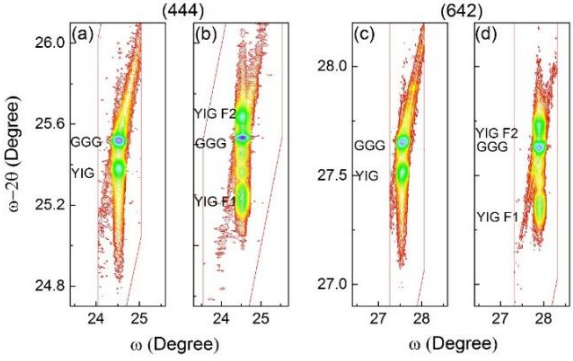

Figure 4: Reciprocal Space Mapping of 200 nm YIG/GGG(111) film; (a) and (b) represent reciprocal space map for oxygen annealed sample (A) and ambient air annealed sample (F) in (444) symmetric direction respectively. Reciprocal space map of sample A and sample F in (642) asymmetric direction are shown in (c) and (d) respectively.

The X-ray pole-figure for sample A and sample F are shown in fig. 5(a) and 5(b) respectively. The Film and the substrate peaks are sharp and appear at the same position, indicating YIG(111)||GGG(111). The plane (444) at $\psi = 0°$ shows two-fold symmetry, the plane (422) at $\psi = 19.47°$ and (400) at $\psi = 54.74°$ show three-fold symmetry, whereas the plane (642) at $\psi = 22.20°$, (640) at $\psi = 36.72°$ and (420) at $\psi = 39.23°$ show six-fold symmetry. This symmetry argument confirms the epitaxial growth.

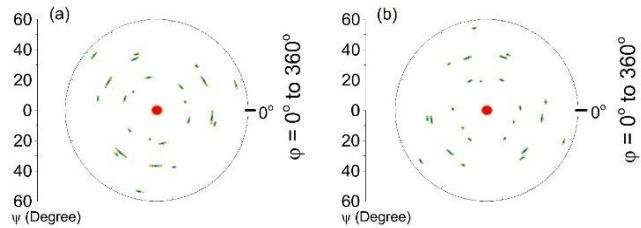

Figure 5: Pole figure; (a) and (b) are representing $\phi$ scan in (444), (422), (642), (640), (420) and (400) planes of the YIG film and GGG substrate for sample A and sample F respectively. The peaks for layer and substrate are indistinguishable due to epitaxial growth.

The differences in the thermal expansion coefficients of the film and substrate combined with thermal cycling during growth and annealing tend to generate non-zero stress in the films.[35,36] The average biaxial stress in a film can be quickly determined from d vs. $\sin^2\psi$ linear fit. The $\theta$-$2\theta$ diffraction scan was performed over different planes of the sample by side angle $\psi$ tilt. The elastic strain of crystal lattice causes shift in the diffraction peak positions which can be detected for each $\psi$ tilt. The strain in crystalline films can be defined as the difference in d-spacing of stressed and unstressed lattice. We performed residual stress measurement on sample A and sample F by using multiple tilt "$sin^2\psi$" technique.[25-27] Panalytical X'Pert Pro four-circle diffractometer was used to measure the stress in both the samples. The strain formula used to derive the fundamental X-ray residual stress equation is given below.

$$\varepsilon_{\phi\psi} = \frac{d_{\phi\psi} - d_0}{d_0} = \frac{1+\nu}{E} \sigma_\phi \sin^2\psi - \frac{\nu}{E}(\sigma_{11} + \sigma_{22})$$

$$\sigma_\phi = \sigma_{11}\cos^2\phi + \sigma_{12}\sin2\phi + \sigma_{22}\sin^2\phi$$

The parameters involved are defined as, $d_0$ - unstressed lattice plane spacing, $d_{\phi\psi}$ - change in lattice plane spacing with $\psi$ tilt and rotated by $\phi$ within the film plane, $\nu$ - poisson's ratio, E - elastic constant, and $\sigma_\phi$ – normal stress in the $\phi$ direction. This equation gives a linear variation of d vs. $\sin^2\psi$ and the stress in the $\phi$ direction may be calculated, provided the values of elastic constant E, unstressed plane spacing $d_0$ and the Poisson's ratio $\nu$ are known.

The strain ($\varepsilon_{\phi\psi}$) or $(d_{\psi\phi} - d_0)/d_0$ vs. $\sin^2\psi$ plot has been shown in figure 6, where the sample A and sample F show linear behaviour. Figure 6(a) represents $(d_{\psi\phi} - d_0)/d_0$ vs. $\sin^2\psi$ plot for sample A. Whereas, $(d_{\psi\phi} - d_0)/d_0$ vs. $\sin^2\psi$ plot for F1 and F2 layers of sample F are shown in figure 6(b) and 6(c) respectively. The reported Poisson's ratio $\nu$ for YIG are 0.29, 0.29 and 0.277 units respectively.[28,37,38] We've used these values to calculate the stress in both the post-treated samples. The elastic modulus E reported by references 38 and 39 are 192.61 GPa[38] and 206 GPa[39] respectively. The elastic Modulus for the third cycle of thermal shock for YIG specimen saturates at a value of ~ 191 GPa.[40] We set an upper and lower bound for Poisson's ratio $\nu$ as 0.277 and 0.29 respectively. Similarly, for elastic modulus E, the upper and lower bounds were set in a range from 190 to 210 GPa. The stress was calculated by fitting $(d_{\psi\phi} - d_0)/d_0$ vs. $\sin^2\psi$ linearly.

*e-mail: rcb@iitk.ac.in

The calculated stress value for oxygen annealed YIG film is -2.157 GPa. Whereas, for air annealed

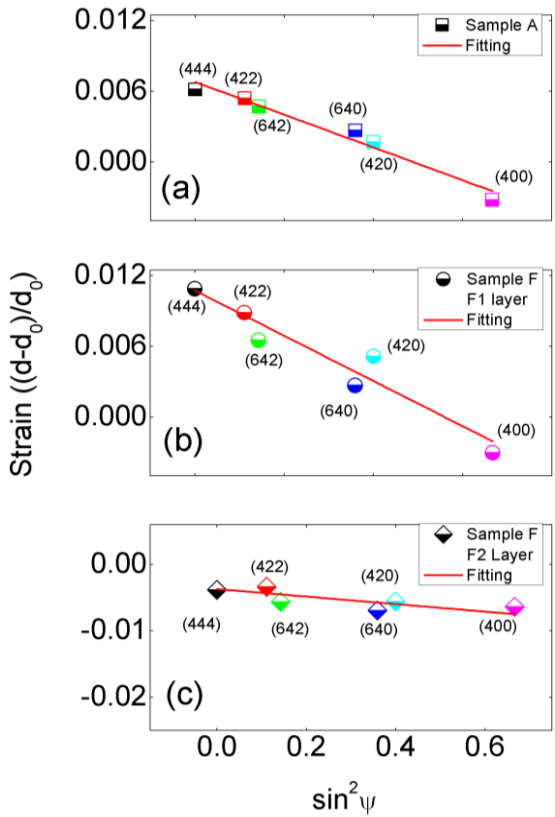

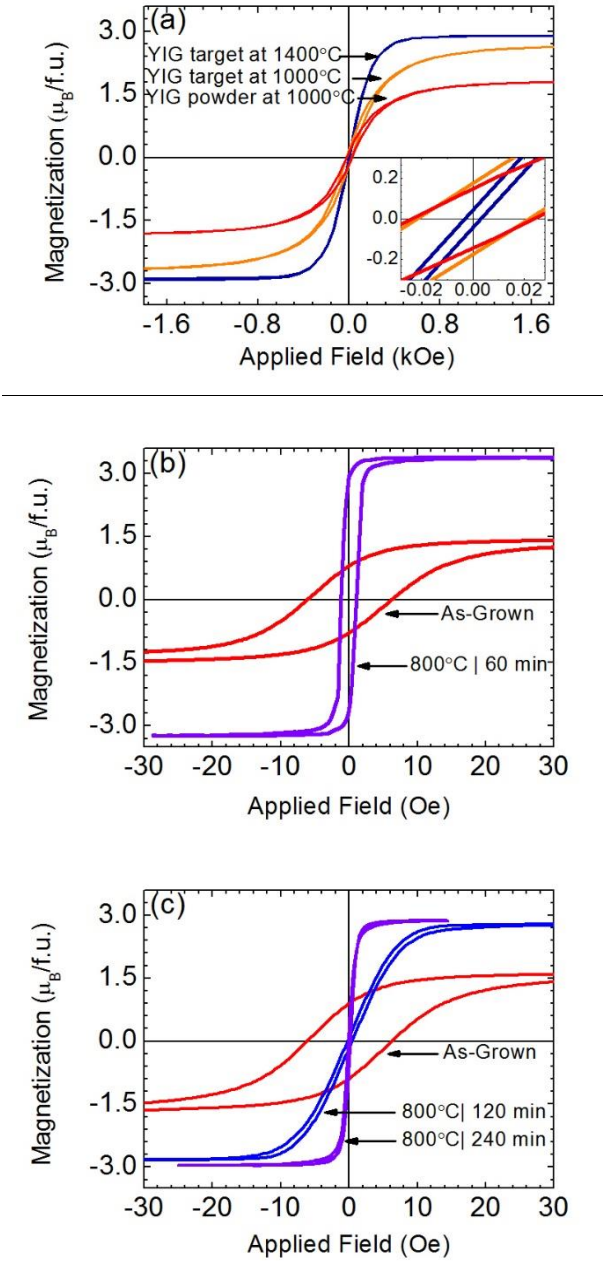

Figure 6: Residual stress measurement using $\sin^2\psi$ technique; (a) $\varepsilon_{\phi\psi} = \frac{d_{\phi\psi} - d_0}{d_0}$ vs. $\sin^2\psi$ for oxygen annealed (A) sample. The plots (b) and (c) represent $\frac{d_{\phi\psi} - d_0}{d_0}$ vs. $\sin^2\psi$ linear plot for ambient air annealed (F) sample's F1 and F2 layers respectively.

YIG's F1 and F2 layers are -2.983 GPa and -0.843 GPa respectively. The negative sign here indicates a state of compressive stress in both the samples. Since the strain in a film relaxed with increasing thickness, and the thickness of the oxygen annealed sample (200 nm) is larger than the thickness of the lower layer (F1) of air annealed sample, the measured stress value for the oxygen annealed sample A is less compare to air annealed sample's lower layer (F1). Whereas, the upper layer (F2) in air annealed sample has almost negligible stress which suggests complete relaxation of the layer to cubic structure. This measurement further confirms the double layer structure with a largely distorted YIG layer in between GGG substrate and a relaxed top YIG layer in air annealed samples.

Figure 7: Room temperature Vibrating Sample Magnetometry (VSM) data using QUANTUM DESIGN Physical Property Measurement System (PPMS); (a) YIG Powder annealed at 1000°C, YIG solid annealed at 1000°C and YIG solid annealed at 1400°C. Inset shows the magnified version. (b) Magnetization comparison of 200nm YIG/GGG(111) film oxygen annealed at 800°C for 60 minutes with reference to the as-grown microcrystalline film. (c) Magnetization of 200 nm YIG/GGG(111) samples ambient air annealed at 800°C for 120 and 240 minutes in air.

We have measured the magnetization of well-crystallized powders and pellets of YIG at room temperature as function of applied magnetic field.

*e-mail: rcb@iitk.ac.in

These data are presented in fig. 7(a). A magnified view of the data near origin is shown in the inset, from which we extract the value of coercive field (H$_C$). The saturation magnetization M$_S$ and coercive fields H$_C$ of the YIG powder at 1000°C, YIG solid pellet annealed at 1000°C and the final YIG solid pellet annealed at 1400°C are 1.88 µ$_B$/f.u. | 23.99 Oe, 2.76 µ$_B$/f.u. | 21.69 Oe and 2.88 µ$_B$/f.u. | 3.32 Oe respectively. The results of magnetization measurements on oxygen annealed samples are shown in Figure 7(b) and the extracted values for saturation magnetization and coercive field are 1.48 µ$_B$/f.u.|6.22 Oe and 3.31 µ$_B$/f.u.|1.12 Oe for as-grown (microcrystalline) and 800°C | 60 mins annealed samples respectively. Figure 7(c) represents the magnetization of slowly cooled air annealed samples at 800°C for 120 minutes and 240 minutes and the measured saturation magnetization and coercive field values are 2.84 µ$_B$/f.u.|0.25 Oe and 2.92 µ$_B$/f.u.|0.12 Oe respectively. The air annealing for larger time period doesn't show any significant enhancement in the saturation magnetization as well coercive field. Setsuo Yamamoto et al.[14] have reported the saturation magnetization (Ms) of reactive RF magnetron sputtered YIG/GGG(111) post-annealed sample at 850°C as ~ 1.30 kG and coercivity less than 5 Oe, whereas in our oxygen annealed sample these are 1.63 kG and ~ 1.12 Oe respectively. Saturation magnetization of PLD grown YIG/GGG(111) sample, annealed at 1000°C in 1 atm oxygen is 2 µ$_B$/f.u.,[22] which is less compare to 3.31 µ$_B$/f.u. for oxygen annealed sample at 800°C in our report. Dorsey et al.[41] have reported magnetization measurements on the YIG film pulsed laser deposited on GGG(111) substrate at 750°C and then cooled in ~1 atm oxygen partial pressure. While their magnetization (1.75 kG) is marginally higher than the value (1.63 kG) for our oxygen annealed sample (sample A) but the minimum coercivity they achieved is ~5 Oe which is much larger than 1.12 Oe in our sample. Further, a saturation magnetization of ~3 µ$_B$/f.u. and coercivity (H$_C$) of 47Oe of PLD grown YIG/GGG(111) was reported by Krockenberger et al.[23] One more report[19] on PLD grown YIG films on GGG(111) lists M$_S$ = 1.25 kG and H$_C$ ~ 1Oe.

We can also calculate the stress-induced magnetic anisotropy field as these pseudomorphic films are coherently strained. The uniaxial anisotropy constant $K_u$ is given by $K_u = K_\sigma + K_c + K_g$, where $K_\sigma$ is stress-induced anisotropy constant, $K_c$ is cubic anisotropy constant and $K_g$ is growth-induced anisotropy constant. Among them, the stress-induced anisotropy $K_\sigma$ is the main contributing term to the total anisotropy for PLD-grown films.[42-44] The stress-induced magnetic anisotropy along the <111> direction of cubic crystals is described as $K_u \sim K_\sigma = -3/2\, \lambda_{111}\sigma_{||}$ and the stress-induced magnetic anisotropy field as $H = -3\lambda_{111}\sigma_{||}/M_s$.[36,42,45] Where, $\sigma_{||}$, $\lambda_{111}$ and $M_s$ are the in-plane stress, magnetostriction constant and saturation magnetization, respectively. The reported $\lambda_{111}$ value for epitaxial YIG/GGG(111) sample is $-1.7 \times 10^{-6}$.[46] As, the ambient air annealed sample has a very thin strained layer at the interface and a thick relaxed top layer. We can assume that the contribution to magnetization is merely due to this bulk like relaxed YIG layer and hence can take a magnetostriction constant value $\lambda_{111} = -3.0 \times 10^{-6}$ of bulk.[47-49] The stress-induced magnetic anisotropy constant and magnetic anisotropy field for sample A and sample F are $K_\sigma = -5.50 \times 10^4\, erg/cm^3$, $H = 0.848\, kOe$ and $K_\sigma = -3.79 \times 10^4\, erg/cm^3, H_\sigma = 0.627\, kOe,$ respectively.

**CONCLUSIONS:**

In summary, we have observed that the PLD as-grown films of YIG on (111) GGG crystals at 800°C in 4E-2 mbar O$_2$ pressure are poorly crystalline and dark brown in physical appearance. The post-deposition annealing, both in full oxygen and in ambient air, recrystallizes these films. The high rate of cooling (10°C/min) after ambient air annealing promotes epitaxial growth, whereas the slow cooling (2°C/min) induces a double layer structure. This latter class of films show Laue oscillations in the X-ray diffraction pattern, a highly textured pole-figure and very low values of FWHM of ω scans. Our structure analysis suggests these equilibrium single crystal like features originate from the top layer, which is much thicker. The saturation magnetization of the oxygen annealed sample is marginally higher than that of the air annealed sample, but both are comparable to bulk value (~ 3 µ$_B$/f.u.) for YIG. However, the air annealed sample has a very low $H_C$ (~ 0.12 Oe) compare to the $H_C$ of the oxygen annealed sample (~ 1.1 Oe). Reciprocal space mapping in symmetric and asymmetric directions for both types of films show that in-plane lattice of the YIG film is locked to the in-plane lattice of the substrate, indicating epitaxial growth. This locking with substrate breaks the cubic symmetry by introducing tetragonal distortion that generates strain in the film. The stress calculation by $\sin^2\psi$ technique yielded qualitative information on the residual stress in both the samples. Our studies have shown that optimization of post-deposition annealing protocols is necessary to achieve high quality thin epitaxial films of this technologically important ferrimagnet.

*e-mail: rcb@iitk.ac.in

**ACKNOWLEDGEMENTS:**

Ravinder Kumar wish to acknowledge the financial support from Indian Institute of Technology (IIT) Kanpur and Prof. R. C. Budhani to the J. C. Bose National Fellowship of the Department of Science and Technology (DST), Government of INDIA. The authors are thankful to Dr. Shubhankar Das, Mr P. C. Joshi, Mr Pramod Ghising, Dr. Himanshu Pandey, Dr. Dushyant Kumar and Dr. Biswanath Samantaray for technical support and fruitful discussion.

1. S. Geller and M. A. Gilleo, J. Phys. Chem. Solids **3**, 30 (1957).
2. F. Bertaut and F. forrat, C. R. Acad. Sci. Paris **242**, 382 (1956).
3. S. Geller and M. A. Gilleo, Acta Cryst. **10**, 239 (1957).
4. V. Cherepanov, I. Kolokolov, V. L'vov, Phys. Rep. **229**, 81 (1993).
5. A. D. Block, P. Dulal, B. J. H. Stadler and N. C. A. Seaton, IEEE Photonics **6**, 0600308 (2014).
6. S. Li, W. Zhang, J. Ding, J. E. Pearson, V. Novosad and A. Hoffmann, Nanoscale **8**, 388 (2016).
7. R. C. Lecraw, E. G. Spencer and C. S. Porter, Phys. Rev. **110**, 1311 (1958).
8. A. Sposito, T. C. M. Smith, G. B. G. Stenning, P. A. J. de Groot, and R. W. Eason, Opt. Mater. Express **3**, 624 (2013).
9. A. Kehlberger, K. Richter, M. C. Onbasli, G. Jakob, D. H. Kim, T. Goto, C. A. Ross, G. Götz, G. Reiss, T. Kuschel, and M. Kläui, Phys. Rev. Appl. **4**, 014008 (2015).
10. B. M. Howe, S. Emori, H. M. Jeon, T. M. Oxholm, J. G. Jones, K. Mahalingam, Y. Zhuang, N. X. Sun, and G. J. Brown, IEEE Magn. Lett. **6**, 3500504 (2015).
11. A. A. Serga, A. V. Chumak and B. Hillebrands, J. Phys. D: Appl. Phys. **43**, 264002 (2010)
12. H. Yokoi, T. Mizumoto, N. Shinjo, N. Futakuchi, and Y. Nakano, Appl. Opt. **39**, 6158 (2000).
13. G. E. W. Bauer, E. S. and B. J. V. Wees, Nature Materials **11**, 391 (2012).
14. S. Yamamoto, H. Kuniki, H. Kurisu, M. Matsuura and P. Jang, Phys. Stat. sol. (a) **201**, 1810 (2004).
15. H. M. Chou and E. D. Case, Materials Science and Engineering **100,** 7 (1988).
16. P. Gerard, Thin Solid Films **114**, 3 (1984).
17. H. Wang, C. Du, P. C. Hammel, and F. Yang, Phys. Rev. B **89**, 134404 (2014).
18. S. I. Olikhovskii, O. S. Skakunova, V. B. Molodkin, E. G. Len, B. K. Ostafiychuk and V. M. Pylypiv, Proceedings of the International Conference Nanomaterials: Applications and Properties **4**, 01PCSI05(4pp) (2015).
19. N. S. Sokolov, V. V. Fedorov, A. M. Korovin, S. M. Suturin, D. A. Baranov, S. V. Gastev, B. B. Krichevtsov, K. Yu. Maksimova, A. I. Grunin, V. E. Bursian, L. V. Lutsev and M. Tabuchi, J. Appl. Phys. **119**, 023903 (2016).
20. P. Ghising, Z. Hossain, and R. C. Budhani, Appl. Phys. Lett. **110**, 012406 (2017).
21. Y. M. Kang, S. H. Wee, S. I. Baik, S. G. Min, S. C. Yu, S. H. Moon, Y. W. Kim and S. I. Yoo, J. Appl. Phys. **97**, 10A319 (2005).
22. Y. Krockenberger, H. Matsui, T. Hasegawa, M. Kawasaki and Y. Tokura, Appl. Phys. Lett. **93**, 092505 (2008).
23. Y. Krockenberger, K.-S. Yun, T. Hatano, S. Arisawa, M. Kawasaki and Y. Tokura, J. Appl. Phys. **106**, 123911 (2009).
24. C. Tang, M. Aldosary, Z. Jiang, H. Chang, B. Madon, K. Chan, M. Wu, J. E. Garay, and J. Shi, Appl. Phys. Lett. **108**, 102403(2016).
25. B. D. Cullity, Addison-wesley Publishing Company, Inc. 1956, chapter 17, p. 431.
26. M. E. Hilley, Ed., Residual stress measurement by X-ray Diffraction, SAE J784a, Society of automotive engineering, Warrendale, PA, 1971, p. 20.
27. I. C. Noyan and J. B. Cohen, B. Ischner and N. J. Grant, Residual Stress: Measurement by Diffraction and Interpretation, Springer-Verlag New York Inc. 1987, p. 122.
28. S. Mino, A. Tate, T. Uno, T. Shintaku and A. Shibukawa, Jpn. J. Appl. Phys. **32**, 3154 (1993).
29. E. L. Jakubisova, S. Visnovsky, H. Chang, and M. Wu, Appl. Phys. Lett. **108**, 082403 (2016).
30. O. G. Ramer and C. H. Wilts, Phys. Status Solidi (b) **79**, 313 (1977).
31. K. P. Belov and A. V. Ped'ko, Sov. Phys. Journal of Experimental and Theoretical Physics **12**, 666 (1961).
32. A. V. Ped'ko, Sov. Phys. Journal of Experimental and Theoretical Physics **14**, 505 (1962).
33. J. P. Hanton and A. H. Morrish, J. Appl. Phys. **36**, 1007 (1965).
34. B. Luthi, Phys. Rev. **148**, 519 (1966).

*e-mail: rcb@iitk.ac.in

35. M. Y. Chern and J. S. Liaw, Jpn. J. Appl. Phys. **36**, 1049 (1997).
36. P. Sellappan, C. Tang, J. Shi and J. E. Garay, Mater. Res. Lett. **5**, 41 (2017).
37. P. J. besser, J. E. Mee, P. E. Elkins and D M Heinz, Mater. Res. Bull. **6**, 1111 (1971).
38. H. M. Chou and E. D. Case, Journal of Materials Science Letters **7**, 1217 (1988).
39. D. F. Gibbons, V. G. Chirba, Phys. Rev. **110**, 770 (1958).
40. H. M. Chou and E. D. Case, Materials Science and Engineering **100**, 7 (1988).
41. P.C. Dorsey, S. E. Bushnell, R. G. Seed and C. Vittoria, J. Appl. Phys. **74**, 1242 (1993).
42. E. A. Giess, B. A. Calhoun, E. Klokhom, T. R. McGuire, and L. L. Rosier, Mater. Res. Bull. **6**, 317 (1971).
43. M. Kubota, A. Tsukazaki, F. Kagawa, K. Shibuya, Y. Tokunaga, M. Kawasaki, and Y. Tokura, Appl. Phys. Exp. **5**, 103002 (2012).
44. H. Pandey, P. K. Rout, Anupam, P. C. Joshi, Z. Hossain and R. C. Budhani, Appl. Phys. Lett. **104**, 022402 (2014).
45. D. M. Heinz, P. J. Besser, J. M. Owens, J. E. Mee, and G. R. Pulliam, J Appl Phys. **42**, 1243 (1971).
46. J. Mada and K. Yamaguchi, J. Appl Phys **1**, 53 (1982).
47. G. P. Vella-Coleiro, Rev. Sci. Instrum. **50**, 1130 (1979).
48. E. R. Callen, A. K, Clark, 8, DeSavage, and W. Coleman, Phys. Rev. **130**, 1735 (1963).
49. P. J. Flanders, R. F. Pearson, and J. L. Page, Brit. J**.** Appl. Phys. **17**, 839 (1966).

*e-mail: rcb@iitk.ac.in